\documentclass[12pt,letterpaper]{article}

\usepackage{parskip} 
\usepackage{graphicx}

\usepackage[utf8]{inputenc}
\usepackage{url,hyperref}
\usepackage{color} 
\usepackage{units}

\usepackage[section]{placeins}

\providecommand{\keywords}[1]{\textbf{\textit{Keywords:}} #1}
\providecommand{\pacs}[1]{\textbf{\textit{PACS:}} #1}

\begin{document}
\onecolumn

\title{Human fields and their impact on brain waves \\ A pilot study} 
\author{Jesus Acosta-Elias$^{1 \ast}$, Santiago Mendez-Moreno$^{1}$,\\ Omar Vital-Ochoa$^{2}$  and Ricardo Espinosa-Tanguma$^{3}$.\\ \\
$^{1}$ Science School of the  Autonomous University of  San Luis Potos\'i\\
$^{2}$ Engineering School of the  Autonomous University of San Luis Potos\'i \\
$^{3}$ Medicine School of the  Autonomous University of San Luis Potos\'i\\
$^{\ast}$ Corresponding author: Jesus Acosta-Elias, jacosta@uaslp.mx\\}

\maketitle
\begin{abstract}
During brain function, groups of neurons fire synchronously. When these groups are large enough, the resulting electrical signals can be measured on the scalp using Electroencephalography (EEG). The amplitude of these signals can be significant depending on the size and synchronization of the neural activity. EEG waves exhibit distinct patterns based on the brain's state, such as whether it is asleep, awake, engaged in mental calculations, or performing other cognitive functions. Additionally, these patterns can be modified by external factors, such as transcranial magnetic stimulation (TMS). TMS involves bringing an antenna that generates variable electromagnetic fields close to specific areas of the skull to treat certain pathologies. Given that the human body naturally generates magnetic fields, a question arises: Can these fields influence the EEG by modulating neuronal function, causing a resonance effect, or through some unknown interaction? This study investigated whether approaching the palm of the hand to the top of the head (Intervention) could induce effects in the EEG. Power Spectral Density (PSD) was obtained for the 30 seconds preceding the intervention (PSD\_pre) and the final 30 seconds of the intervention (PSD\_last). The exact Wilcoxon signed-rank test suggests that the median of PSD\_pre is greater than the median of PSD\_last at the 95\% confidence level (p-value = 0.004353). In contrast, in the control group, the test indicates that at the 95\% confidence level (p-value = 0.7667), the median of PSD\_pre is not greater than the median of PSD\_last.
\end{abstract}
 
\\

\section{Introduction}
Electroencephalography (EEG) is a non-invasive technique used to measure and record the electrical activity of the brain. It provides insights into neural function by detecting voltage fluctuations generated by ionic current flows within neurons. EEG records brain activity via electrodes placed on the scalp. These electrodes detect the summated activity of synchronized neuronal populations, particularly postsynaptic potentials in the cortical pyramidal neurons\cite{Leif}.

Physical and mental activities, such as moving the tongue or performing a mental arithmetic operations, can generate artifacts that alter the EEG signal.  Artifacts can also be caused by external factors like the 50 or 60 Hz power supply frequency. Additionally, certain pathologies can modify these signals. Therefore, the study of EEG signals has become a very useful tool in neurology, psychiatry, psychology, and others fields. The EEG can also be influenced by electric and magnetic fields, such as when the brain is subjected to transcranial magnetic stimulation \cite{deepBrainA, deepBrainB}. 

It is known that the human body naturally exhibits magnetic fields \cite{biomagnetismoA, biomagnetismoB}, and it is also thought that some brain functions, such as consciousness, may result from quantum processes \cite{quantumA, quantumBrain}. Additionally, it is suspected that new physics may be hidden in living matter \cite{livingMatter}, suggesting that there may be unknow laws of physics and that in living matter they could become visible. This raises the question: Can these fields, not necessarily magnetic and with properties inherent to the human body (amplitude, frequency, phase) influence the EEG by modulating neuronal function, causing a resonance effect, or through some unknown interaction?

\section{Methodology}
To study whether the electric/magnetic field of the human body can modify EEG signals, the protocol involved the experimenter bringing the palm of the hand closer to the top of the test subject's head without making contact. The EEG were recorded before and during the procedure. This work was conducted in accordance with the principles embodied in the Declaration of Helsinki and local statutory requirements (Aproveed by the Research Committee of the Engineering School of the Autonomous University of San Luis Potos\'i. This Committee reviews  technical and ethical issues). All subjects provided informed consent.

Thirty healthy students, aged 20-24 years, were recruited for the study: twenty for the control group and thirty for the experimental group, inclusive of individuals of all genders. All participants were from the Engineering School of the UASLP. They were interviewed to explain the procedure and assess their health status, primarily to rule out those with any psychological or psychiatric treatment, and to schedule the day and time of the experiment. Subjects were only informed that EEG measurements would be taken for an ongoing study; they were not informed about the procedure in which the researcher would bring their hand close to the top of the subject's head. They were told that the researcher would stand to the side to ensure the electrodes remained connected.

The experiment was conducted by two PhD students (Experimenters) involved in this work. We used the Biopac$^{\tiny{\copyright}}$ system to measure EEG signals by placing three electrodes on the scalp according to the 10-20 system \cite{10-20a}: one electrode at Cz, and the other two at A1 and A2. In the 10-20 system, Cz is the central position located at the intersection of the midline between the nasion and inion and the line connecting the left and right preauricular points on the scalp. The nasion is the indentation between the forehead and the nose, and the inion is the bony protrusion at the base of the skull at the back of the head. A1 and A2 are the electrode positions located on the left and right earlobes, respectively, used as reference or ground electrodes. 

Protocol Description\\
Two protocols were implemented: a control protocol and an intervention protocol. In the control protocol, only EEG readings were taken without any intervention. In the intervention protocol, the experimenter approached their hand to the top of the subject's head and held it there for 60 seconds. The intervention protocol was performed first, followed by the control protocol two weeks later. Some students participated in both protocols, while others did not. The detailed protocols are as follows:

 Control protocol

\begin{enumerate}
 	\item Setup
    		\begin{itemize}
    		\item The subject was seated, and electrodes were placed on their scalp at specific locations (Cz, A1, A2) according to the 10-20 system.
    		\item The subject was instructed to close their eyes and relax naturally, explicitly avoiding yoga or meditation techniques to minimize potential alterations in EEG readings,  while the experimenter remained nearby.
     \end{itemize}
     \item{Data Collection}
     		\begin{itemize}
   		\item EEG data were recorded during two distinct periods: 
   		\begin{itemize}
     			\item First 40 seconds: Baseline relaxation.  
     			\item Next 60 seconds: A control phase where no intervention was applied.  
     	\end{itemize}
    \item At the end, the electrodes were removed, and the subject was thanked for their participation.  
     \end{itemize} 
\end{enumerate}

Intervention protocol
\begin{enumerate}
 	\item Setup
    		\begin{itemize}
    		\item The subject was seated, and electrodes were placed on their scalp at specific locations (Cz, A1, A2) according to the 10-20 system.
    		\item{The subject was instructed to close their eyes and relax naturally, explicitly avoiding yoga or meditation techniques to minimize potential alterations in EEG readings,  while the experimenter remained nearby.}
     \end{itemize}
     \item{Data Collection}
     		\begin{itemize}
   		\item EEG data were recorded during two distinct periods: 
   		\begin{itemize}
     			\item First 40 seconds: Baseline relaxation.  
     			\item Next 60 seconds: The experimenter slowly moved their hand toward the subject's scalp (position Cz) and kept it there during this period.     			
     		    \end{itemize}
     			\item At the end, the electrodes were removed, and the subject was thanked for their participation.  
     \end{itemize} 
\end{enumerate}

Data Acquisition and Processing

\begin{itemize}
\item Equipment: Biopac Student Lab$^{\tiny{\copyright}}$ software, version 4.1 (BSL) and MP36$^{\tiny{\copyright}}$ hardware were used to record EEG data.
\item Data Analysis: Using the BSL software, for the two groups, control and intervention,  the Power Spectral Density (PSD) was calculated for the first 30 seconds of the baseline relaxation (PSD\_pre) and the final 30 seconds of the intervention (PSD\_last). These PSD values were derived from the raw EEG signal rather than being calculated separately for the individual EEG bands (delta, theta, alpha, beta, and gamma). This approach was chosen because changes in EEG activity might occur in only one or two bands for each subject, while the other bands remain unaffected. Analyzing each band separately could result in a lack of statistical significance during hypothesis testing due to this variability. By analyzing the PSD from the raw signal, any modifications; whether in the delta band for one subject or the gamma band for another are captured comprehensively in the analysis.

\item Statistical analysis: The Shapiro-Wilk normality test indicated that the data (PSD\_pre and PSD\_last) did not follow a normal distribution in both the control and intervention groups. Consequently, the Wilcoxon signed-rank test was used for paired data analysis. This non-parametric test was chosen as it does not assume normality and is suitable for comparing two related samples. The analysis was conducted using the R software, leveraging its statistical tools for hypothesis testing.

\end{itemize}

\section{Results}

The results obtained indicate, with a 95\% confidence level $(p-value = 0.004353)$, that when the experimenter's hand is positioned above the participants' heads without physical contact and without informing them of this action, the PSD\_pre is significantly higher than the PSD\_last.

\begin{figure}[!ht]
	\begin{center}

		\includegraphics[origin=c,angle=-0, width=14cm]{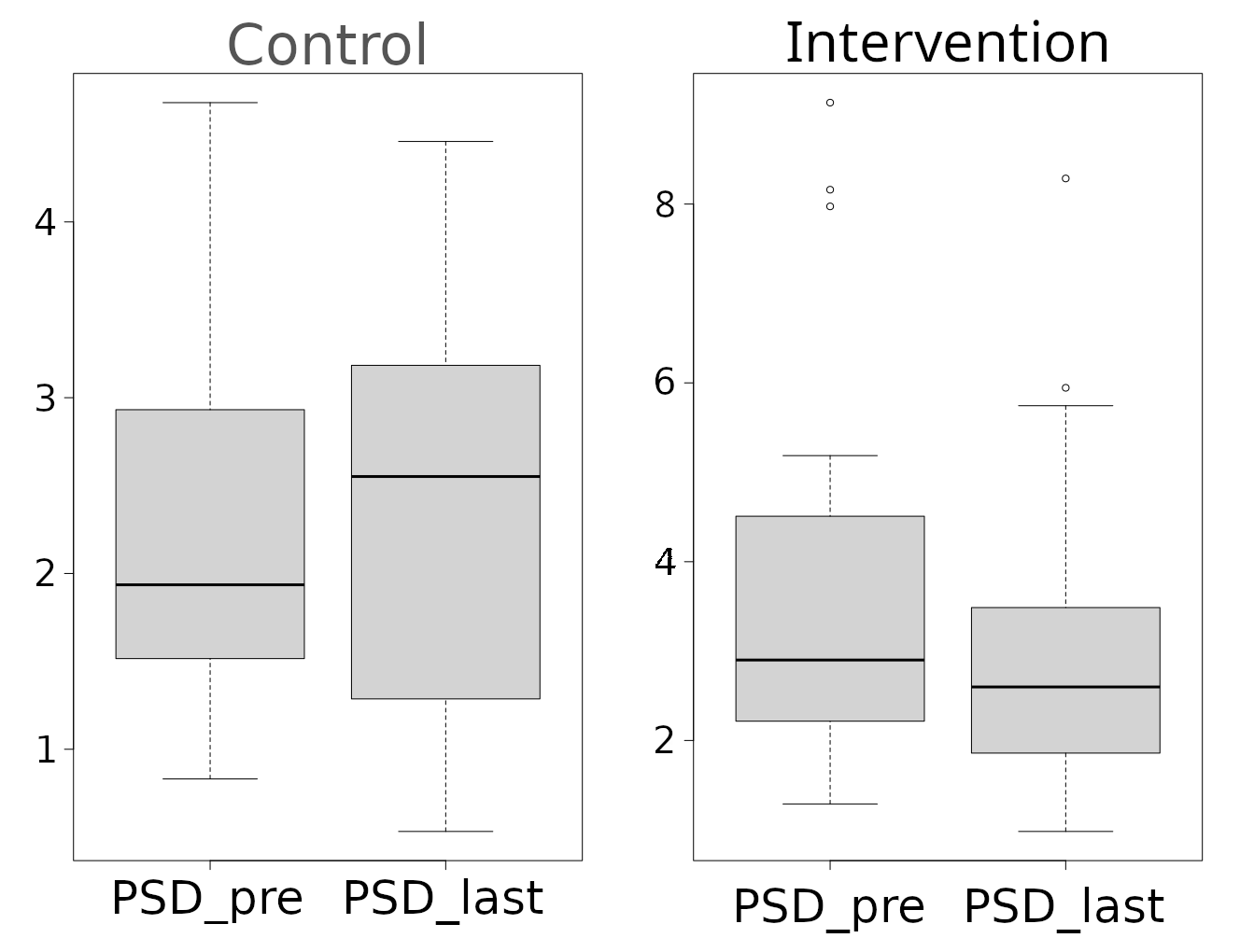}
		\caption{It can be observed that in the control group, PSD\_pre is lower than PSD\_last. Conversely, in the intervention group, PSD\_pre is higher than PSD\_last. It can also be observed that the PSD\_last of the control group is higher than the PSD\_last of the intervention group.  }
		\label{f1}
	\end{center}
\end{figure}

In the control group, the opposite occurs: the test indicates that at the 95\% confidence level $(p-value = 0.7667)$, the median of PSD\_pre is not greater than the median of PSD\_last. In \cite{PSDadormilado}, Timothy Howarth and colleagues observed that EEG Power PSD is higher in individuals with excessive daytime sleepiness compared to those without. This increase is attributed to a greater presence of the delta band, which is characteristic of deep sleep \cite{Leif}. Based on this, it is reasonable to speculate that the PSD in the control group of the protocol presented here may have increased due to the subjects transitioning from a fully awake state to a drowsy state. This transition could be attributed to spending 100 seconds (40 seconds in the Baseline relaxation state followed by 60 seconds in the control period) with their eyes closed in a relaxed state, which may have facilitated the onset of drowsiness or sleepiness.

\section{Discussion}

This work has several shortcomings. First, all the test subjects (Intervention and control) are undergraduate students, meaning they represent a very specific segment of the population. Second, although the results appear statistically significant, the sample size is relatively small, comprising only 30 subjects in the intervention group and twenty in the control group. Finally, and perhaps most importantly, we cannot control the thoughts and feelings of the subjects during the experiment, which may introduce a significant source of variability. 

Care was taken to ensure that the test subjects were not under any medical treatment of any kind. However, several records had to be excluded because they were classified as outliers, with values far beyond the expected range. When interviewing the outlier subjects, two reported having recently experienced the loss of a loved one. Another said his girlfriend had recently broken up with him.

It remains unclear whether the psychological states reported by these outliers caused the substantial increases in the amplitude of their EEG readings or if this was merely coincidental (see figure 2). Moreover, other outliers reported no psychological conditions deviating from the norm, yet their EEG-derived PSD values were far outside the expected range.

\begin{figure}[!ht]
	\begin{center}

		\includegraphics[origin=c,angle=-0, width=12cm]{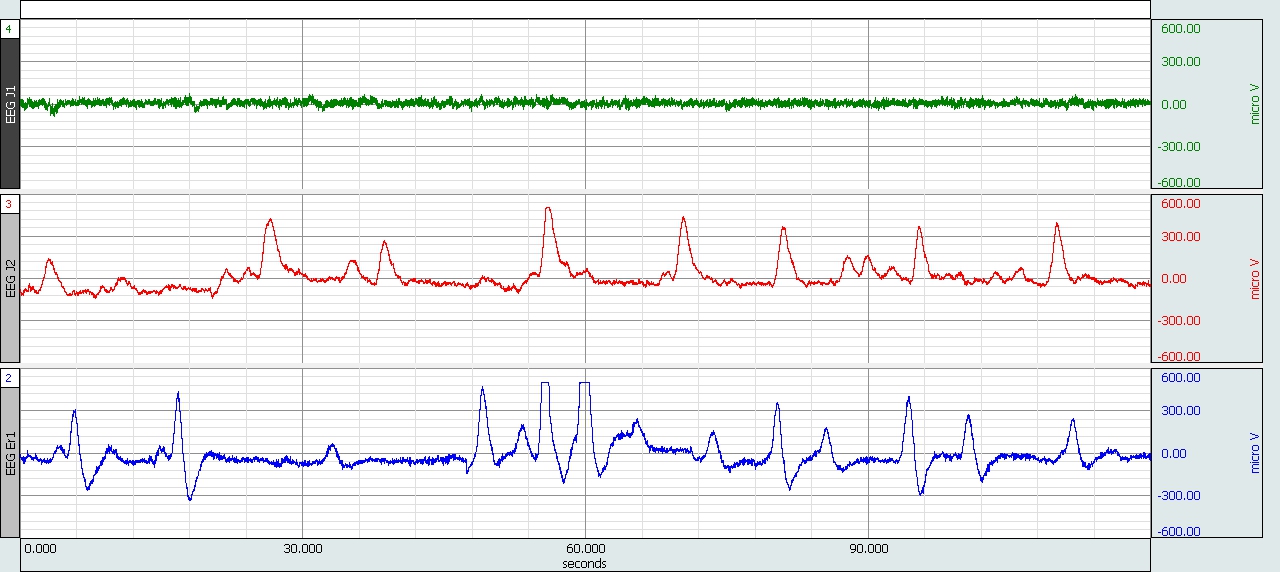}
		\caption{This figure shows a normal EEG reading and two readings considered to be out of range. The green line (PSD\_pre =  5.18661 $\nicefrac {\mu V^{2}}{Hz}$, PSD\_last = 3.17072 $\nicefrac {\mu V^{2}}{Hz}$) represents an EEG obtained using the intervention protocol. This reading, which we can consider normal, corresponds to a 21-year-old female student. The red line (PSD\_pre = 43.03761 $\nicefrac {\mu V^{2}}{Hz}$, PSD\_last = 91.12766 $\nicefrac {\mu V^{2}}{Hz}$) corresponds to the control protocol for the same student, but after experiencing the loss of a very close loved one-the death of her grandfather. The timing of the readings, before and after this significant event, was entirely coincidental. (Note: The intervention protocol was conducted first, followed by the control protocol two weeks later.) We do not know if this change in her EEG is related to the loss she experienced or if it is due to other causes.  The blue line (PSD\_pre = 213.35603 $\nicefrac {\mu V^{2}}{Hz}$, PSD\_last = 212.77476 $\nicefrac {\mu V^{2}}{Hz}$) corresponds to the intervention protocol of a 23-year-old student. In an interview after the intervention, we did
not find a possible explanation for his out-of-range readings.}
		\label{outliers}
	\end{center}
\end{figure}

The experiments were conducted with great care. The entire process was reviewed to ensure it was performed correctly. The electrodes had sufficient conductive gel, were well-placed (immobilized using Velcro$^{\tiny{\copyright}}$ tape), and had impedance within tolerance. The experiments took place in a noise-free environment with normal lighting, and various sources of artifacts were avoided as much as possible. However, the experiment is not free of possible artifacts.

\section{Conclusions}

The results seem to indicate that when the palm of the hand was brought close to the Cz point of a group of subjects sitting in a resting state with their eyes closed, and without being informed of the intervention, the PSD during the last thirty seconds of the intervention decreased compared to the first thirty seconds prior to it. In the control group, the opposite occurred: instead of decreasing, the PSD increased. The results are statistically significant at a 95\% confidence level.

Funding statement\\
This work was partially supported by the Autonomous University of San Luis Potos\'i. No additional specific grant funding was received.\\

Availability of Data and Materials\\
The datasets generated and/or analysed during  the current study are available in the repository:\\ https://purl.archive.org/brain-waves/data

Conflict of Interest Statement\\
The authors declare they don’t have any conflict of interest with respect to the subject of this manuscript.

Authors contributions\\
JAE: Conceptual design, Formal Analysis and Writing original draft.\\
SMM: Experiment execution.\\
OVO: Experiment execution.\\
RET: Conceptual design.\\
All authors read and approved the final manuscript\\

\bibliographystyle{unsrt}
\bibliography{brain.bib}
\end{document}